\documentclass[%
prl,
reprint,
footinbib,
 amsmath,amssymb,
 aps
]{revtex4-1}

\usepackage{graphicx}
\usepackage{dcolumn}
\usepackage{bm}
\usepackage{footmisc}
\usepackage{amsmath,amssymb} 
\usepackage{mathrsfs}
\usepackage{amsfonts}
\usepackage{pgfplots}
\pgfplotsset{compat=1.5}
\usepackage{soul}
\usepackage{lipsum}
\usepackage{hyperref}
\hypersetup{
	colorlinks = false,
	linkbordercolor = {orange},
	citebordercolor= {green}
}
\usepackage{floatrow}

\def\IB#1{\boldsymbol{#1}} 
\def\W/!i#1{\Wi} 
\def\bten#1{\IB{\mathsf{#1}}}

\definecolor{Green}{HTML}{006600}
\newcommand{\ac}[1]{{\color{blue} #1}}

\begin{document}


\title{How inertial lift affects the dynamics of a microswimmer in Poiseuille flow}


\author{Akash Choudhary\textsuperscript{1}}
\email{a.choudhary@campus.tu-berlin.de}
\author{Subhechchha Paul\textsuperscript{2}}
\author{Felix R\"uhle\textsuperscript{1}}
\author{Holger Stark\textsuperscript{1}}
\email{holger.stark@tu-berlin.de}

\affiliation{%
	$ {}^{1} $ Institute of Theoretical Physics, Technische Universit\"{a}t Berlin, 10623 Berlin, Germany
}%
\affiliation{%
	$ {}^{2} $ Department of Mechanical Engineering, Indian Institute of Engineering Science and Technology Shibpur 711103, India
}%
%


\begin{abstract}
 We analyze the dynamics of a microswimmer in pressure-driven Poiseuille flow, where fluid inertia is small but non-negligible.
Using perturbation theory and the reciprocal theorem, we show that in addition to the classical inertial lift of passive particles,
the active nature generates a `swimming lift', which we evaluate for neutral and pusher/puller-type swimmers. Accounting 
for fluid inertia engenders a rich spectrum of novel complex dynamics including bistable states, where tumbling coexists with 
stable centerline swimming or swinging. 
The dynamics is sensitive to the swimmer's hydrodynamic signature and goes well beyond the findings at vanishing fluid inertia. Our work will have non-trivial implications on the transport and dispersion of active suspensions in 
microchannels.
%
%
\end{abstract}

\maketitle



\section{Introduction}

Self-propelling microswimmers often experience dynamic fluid environments and confinements, for example, pathogens in lung mucus \cite{levy2014pulmonary}, microorganisms in laminar flow a through porous matrix \cite{bhattacharjee2019bacterial}, 
and sperm cells in the Fallopian tubes \cite{riffell2007sex}.
Often these swimmers interact with micro-scale flows and boundaries \cite{Schaar15}  to enhance survival probability \cite{daddi2021hydrodynamics} and biofilm formation \cite{conrad2018confined}
or cause intriguing collective patterns \cite{Ruehle20}.
Their envisioned artificial counterparts---designed to 
execute \textit{in vitro} drug delivery---also would have to interact with the dynamic conditions of such biological flows \cite{nelson2014micro,zhou2021magnetically}.
Examining the dynamics of microswimmers can help us get insights into the dispersion of active suspensions \cite{peng2020upstream}, it can also  provide guidelines for  the  rational fabrication of microfluidic  drug delivery and for minimizing biofilm formation in biomedical equipments \cite{rusconi2010laminar}.
%

Sheared flows in biological systems and microchannels impose substantial vorticity on the swimmer, which results in continuous tumbling.
Past experimental and computational studies have shown that this tumbling, in conjunction with surface interactions, cause upstream swimming known as rheotaxis \cite{hill2007hydrodynamic,nash2010run,tung2015emergence,Mathijssen19}.
%
%
\citet{zottl2012nonlinear,zottl2013periodic} developed a theoretical model in the
Stokes regime that captured swimming in Poiseuille
flows and they reported upstream swinging and downstream tumbling states in planar and cylindrical channels.
Related experimental and theoretical research studied single swimmer trajectories \cite{Uppaluri12,Junot21},  shear induced trapping \cite{rusconi2014bacterial}, and focussing of phototactic algae \cite{Garcia13} or  magnetotactic bacteria \cite{Waisbord16,Meng18}.
Most recently, \citet{peng2020upstream} investigated Taylor dispersion in active suspensions.

In recent years,
experimental and theoretical studies have shown how inertia affects the unsteady propulsion of ciliated \cite{wang2012unsteady,hamel2011transitions} and larger swimmers \cite{wang2012inertial,khair2014expansions}. 
The influence of particle inertia has been discussed in Refs.\ \cite{low2018inertial,lowen2020inertial}.
With the recent advent of high-speed tunable microswimmers \cite{ren20193d,aghakhani2020acoustically,zhou2021magnetically}, 
understanding the effects of inertia can help in effective designs of biomedical devices.
However, little is known how fluid inertia affects swimmer dynamics in sheared flows, which we will address in this article.

For passive particles the Segr\'e-Silberberg effect at finite Reynolds numbers
has been known for decades \cite{segre1961,segre1962b}.
Inertial lift forces cause cross-stream migration and eventually focus particles roughly halfway between channel center and walls.
This effect has initiated major advances in cell-sorting and flow cytometry techniques in the newly developing field of inertial microfluidics \cite{di2009inertial,zhang2016}.
To understand it, we note that a rigid particle resists the strain in background flow and generates a stresslet disturbance in the fluid decaying as $ 1/r^{2}$
\cite{batchelor1967}.
The disturbance interacts with  the  curvature of the background flow and the channel walls,
which in the presence of fluid inertia results in counter-acting shear-gradient and wall-induced lift forces that cause inertial focusing \cite{ho1974}. 



In his seminal work, \citet{saffman1965} 
considered a particle moving relative to a uniform shear flow under the influence of 
an external (gravitational) 
force. He showed that it also experiences a cross-streamline lift.
Similar investigations
were presented in recent works on electrophoresis \cite{kim2009,yuan2016,choudhary2019inertial,khair2020migration,khair_comment}.
They stressed the key role of 
the leading hydrodynamic multipole generated by the particle. 

Microswimmers also move relative to an applied background flow. In this  article we consider the generic source-dipole and force-dipole 
microswimmers and calculate the resulting swimming lift in a planar Poiseuille flow when fluid inertia is small but non-negligible. 
We demonstrate that, in combination with the passive inertial lift, this gives rise to novel and rich complex dynamics in channel flow,
which goes well beyond the findings in Refs.\ \cite{zottl2012nonlinear,zottl2013periodic}.
Our work thereby opens up a new direction in the field of active matter by connecting research on microswimmers to the field
 of inertial microfluidics with all its biomedical applications \cite{di2009inertial,zhang2016}.



\begin{figure} 
	\centering
	\includegraphics[width=.55\columnwidth]{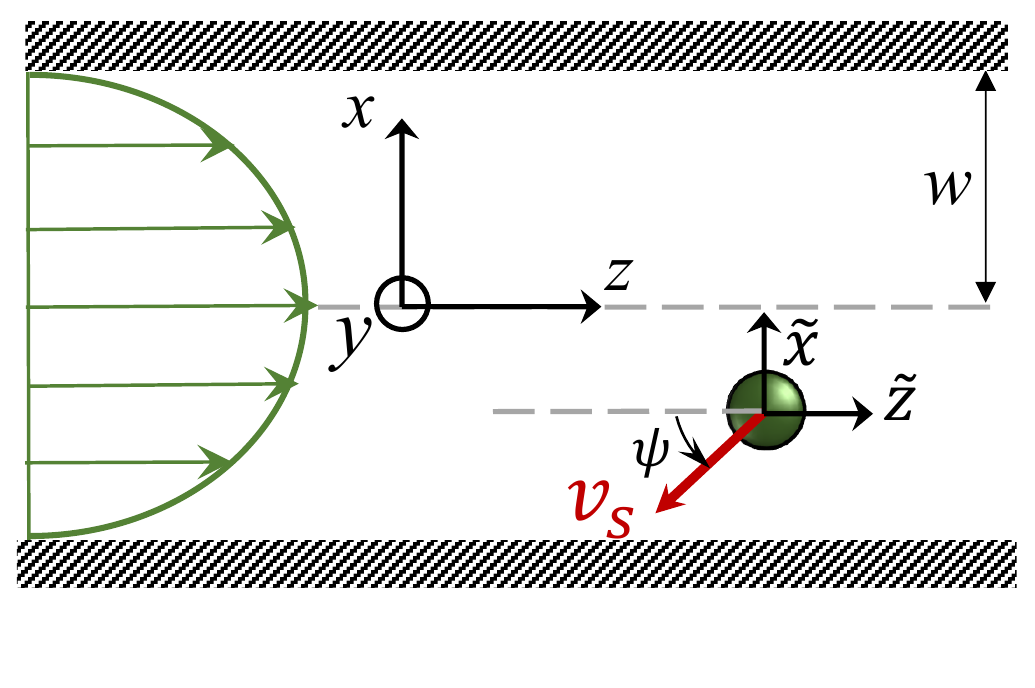}
	\caption{
		A spherical microswimmer with 
		velocity $v_s  \IB{p}$ moves in a planar Poiseuille flow inside a channel with 
		half width $w$. The coordinate frame $\{ \tilde{x},\tilde{y},\tilde{z} \}$ 
		co-moves
		with the swimmer.} 
	\label{fig:schematic}%
\end{figure}

In the following, we consider a spherical swimmer of radius $a$ that self-propels with  velocity $ \IB{v}_{s} = v_s \IB{p}$ in
a two-dimensional Poiseuille flow  $ \IB{v}_{m} = v_{m} [1-({x}/{w})^{2}] \, \IB{e}_{z} $,
where $ v_{m} $  is
the maximum flow velocity and $ w $ the half channel width (see Fig.\ \ref{fig:schematic}).
The overdamped motion of a noise-free swimmer can be described by
dynamic equations for swimmer position ($ \IB{r}$) and orientation ($ \IB{p}$) vector,
\begin{equation}\label{kin_dim}
\IB{\dot{r}}  = \IB{p} +  \bar{\IB{v}}_{m}  +\mathcal{F} (x,\psi) \, \IB{e}_{x}\, , \quad 
\IB{\dot{p}}  = \frac{1}{2} \left(\nabla  \times \bar{\IB{v}}_{m}  \right) \times \IB{p}, 
\end{equation}
where we rescaled velocities by swimming speed $v_s$, lengths by $w$, and time by $ w/v_{s} $. $ \mathcal{F} $  denotes the total inertial lift velocity, which comprises the passive and swimming lift.
It vanishes when fluid inertia becomes negligible and
the system moves in the Stokesian regime as studied in Ref. \cite{zottl2012nonlinear}. 
%
The passive inertial lift is well-explored \cite{ho1974,schonberg1989,asmolov1999inertial} and, except in close vicinity to the channel walls, can be well approximated by
	$ \mathcal{F}_{\text{passive}}(x) \approx \text{Re}_{p} \, \kappa \bar{v}_{m} \, x \left(  1  -  {x^{2}}/{x_{eq}^{2}}   \right) $, as we summarize in the supplemental material. Here,
$ x_{eq} $ denotes the stable equilibrium positions, 
$\kappa = a/2w$ the ratio of swimmer radius to channel width,
and 
$\bar{v}_{m} = v_m/v_s$ is the scaled centerline flow velocity. 
The swimmer Reynolds number $ \text{Re}_{p} = \rho v_{m}  \kappa a/\mu $ is based on the characteristic shear around a swimmer; 
$ \rho$ and $ \mu $ represent the fluid density and viscosity, respectively.

\section{Results and Discussion}

\subsection{Neutral Swimmers}
To evaluate the additional swimming lift $ \mathcal{F}_{swim}$, we 
find the disturbance field $ \IB{v} $ created by the microswimmer using the continuity and the quasi-steady Navier-Stokes equations 
in the co-moving swimmer frame $ \lbrace \tilde{x},\tilde{y}, \tilde{z}  \rbrace $, 
\begin{equation}\label{GE}
	\tilde{\nabla} \cdot \IB{v} = 0 \, , \quad \text{Re}_{p} \, \IB{f}  = \tilde{\nabla} \cdot \bten{\sigma} \, .
\end{equation}
Here, $ \IB{f} =   \IB{v} \cdot \tilde \nabla  \IB{v}_{\infty} + \IB{v}_{\infty} \cdot \tilde \nabla  \IB{v} + \IB{v} \cdot \tilde \nabla  \IB{v} $
results from the convective acceleration
with $\IB{v}_{\infty}$
the Poiseuille flow 
field in the
swimmer frame, and $ \bten{\sigma} = -p \bten{I} + 2\bten{e} $ is the  Newtonian stress tensor of the 
disturbance field, where $p$ and $\bten{e}$ represent pressure and the rate-of-strain tensor, respectively.
First, we consider a neutrally buoyant microswimmer that generates a source-dipole disturbance that, in leading order, resembles the flow field generated by some ciliated microswimmers
\cite{evans2011orientational} and active droplets \cite{thutupalli2011swarming,Schmitt16}.
%
%
Before evaluating the inertial swimming lift, we will perform an order-of-magnitude analysis to predict its scaling for small $\text{Re}_{p}$.
This will provide a fundamental understanding how weak inertia affects 
swimmer motion.

The classical analyses of \citet{oseen1910uber} and \citet{saffman1965} demonstrated that the magnitude of inertial perturbations increases with distance from the swimmer until an asymptotic ``cross-over radius'', beyond which the 
perturbations become singular.
For the current swimmer system the cross-over radius is $ r_{c} \sim \text{Re}_{p}^{-1/2} $ \cite{saffman1965} that divides the 
entire domain
in \textit{inner} (regular) and \textit{outer} (singular) regions. Substitution of $ r_{c} $ in  the hydrodynamic signature of 
a neutral swimmer ($ \sim 1/r^{3} $) suggests that singular lift is inferior to regular lift, \emph{i.e.},
$ \mathcal{F}_{\text{swim}}  \propto \text{Re}_{p} $. 
This is in contrast to the Saffman lift of a forced particle, where the singular contribution $\propto \text{Re}_{p}^{1/2}$ 
dominates. Hence, we implement a regular perturbation expansion, which turns the Navier-Stokes equations (\ref{GE}) into 
Stokes problems of zeroth ($ \tilde{\nabla} \cdot \bten{\sigma}_{0} =0 $) and  first order 
($\tilde{\nabla} \cdot \bten{\sigma}_{1} = \IB{f}_{0} $), as detailed in the supplemental material.
Using the reciprocal theorem, we are able to calculate the swimming lift velocity 
from the first-order problem \cite{ho1974}
\begin{equation}\label{lift}
\mathcal{F}_{\text{swim}}   = - \frac{\text{Re}_{p}}{6\pi} \int_{V} \IB{v}^{t} \cdot \IB{f}_{0} \;  {\rm d}V.
\end{equation}
Here, 
the auxiliary velocity field $ \IB{v}^{t} $ belongs to a forced particle moving along
the $ x-$direction \cite{ho1974}. 
The convective acceleration $ \IB{f}_{0} $  corresponds to the Stokes solution $ \IB{v}_0 $ of the microswimmer consisting of a
 source-dipole field, which we adopt from the squirmer model \cite{Lighthill52,Blake71,zottl2016emergent},
and a stresslet generated by the shearing  background flow with rate-of-strain tensor $ \bten{e}_{\infty}$,
\begin{align}\label{dist}	
\IB{v}_0 =	\frac{v_{s}\IB{p} }{2 \tilde{r}^{3}} \IB{\cdot} \left[    \frac{3 \tilde{\IB{r}} \tilde{\IB{r}} }{\tilde{r}^{2}}   -\bten{I} \right]   
		- \left[  \frac{5  \bten{e}_{\infty}  \IB{:}  \tilde{\IB{r}}  \tilde{\IB{r}} }{2 \tilde{r}^{5}}  \left(\tilde{\IB{r}} -\frac{\tilde{\IB{r}} }{\tilde{r}^{2}}\right) +\frac{\bten{e}_{\infty} \IB{\cdot} \tilde{\IB{r}}}{   \tilde{r}^{5} }
		\right]. \nonumber
\end{align}
Using the corresponding $\IB{f}_0$ in Eq.\ (\ref{lift}) and $\bten{e}_{\infty}$ for the Poiseuille flow, results in the
inertial swimming lift velocity given in units of $v_s$:  $ \mathcal{F}_{\text{swim}}  = - (7/6) \text{Re}_{p}  \,x  \cos \psi $. 
Thus,  the total inertial lift to be used in Eq. (\ref{kin_dim}) becomes
\begin{equation}\label{F}
\mathcal{F}  =  \text{Re}_{p} \left[  \kappa  \bar{v}_{m} 
 x\left( 1-\frac{x^{2}}{x_{eq}^{2}} \right)   -  x  \cos \psi  \right] \, .
\end{equation}
where we skip the factor $7/6$ for simplicity
\footnote{{We also calculated the modification to z-direction swimmer velocity and y-direction rotational velocity. The former is $ -5/18 Re_{p} x \sin{\psi} $ and latter is found to be identically zero at the present order of approximation. 
}}.

\begin{figure}[b] 
	\centering
	\includegraphics[width = 0.7 \columnwidth]{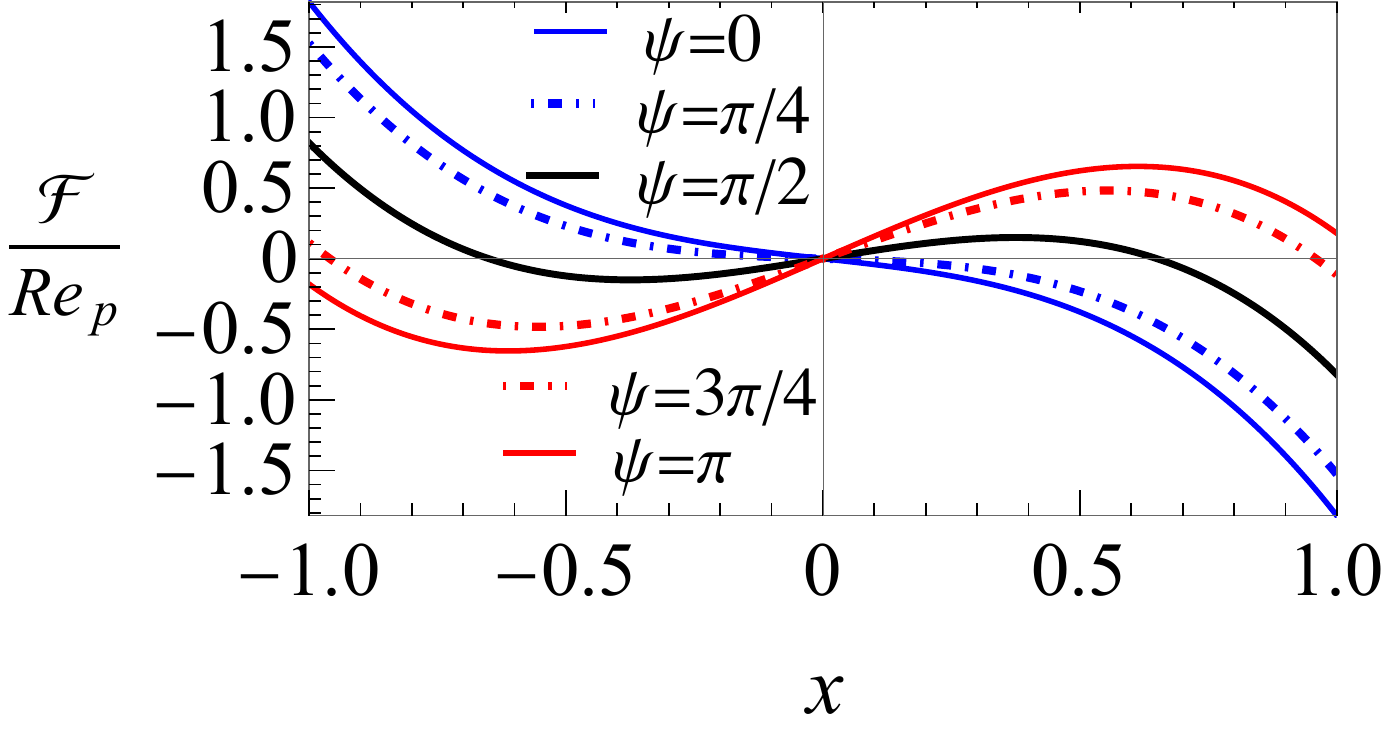}
	\caption{Inertial lift-velocity profiles of a source-dipole 
or neutral microswimmer for different orientation angles 
$ \psi $
	for moderate flow speed $ \bar{v}_{m}  = 6$, $\kappa = 0.1$, and $x_{eq} = \pm 0.65$.
\label{fig:lift_profile}%
}
\end{figure}

\begin{figure}
	\centering
	\includegraphics[scale=0.295]{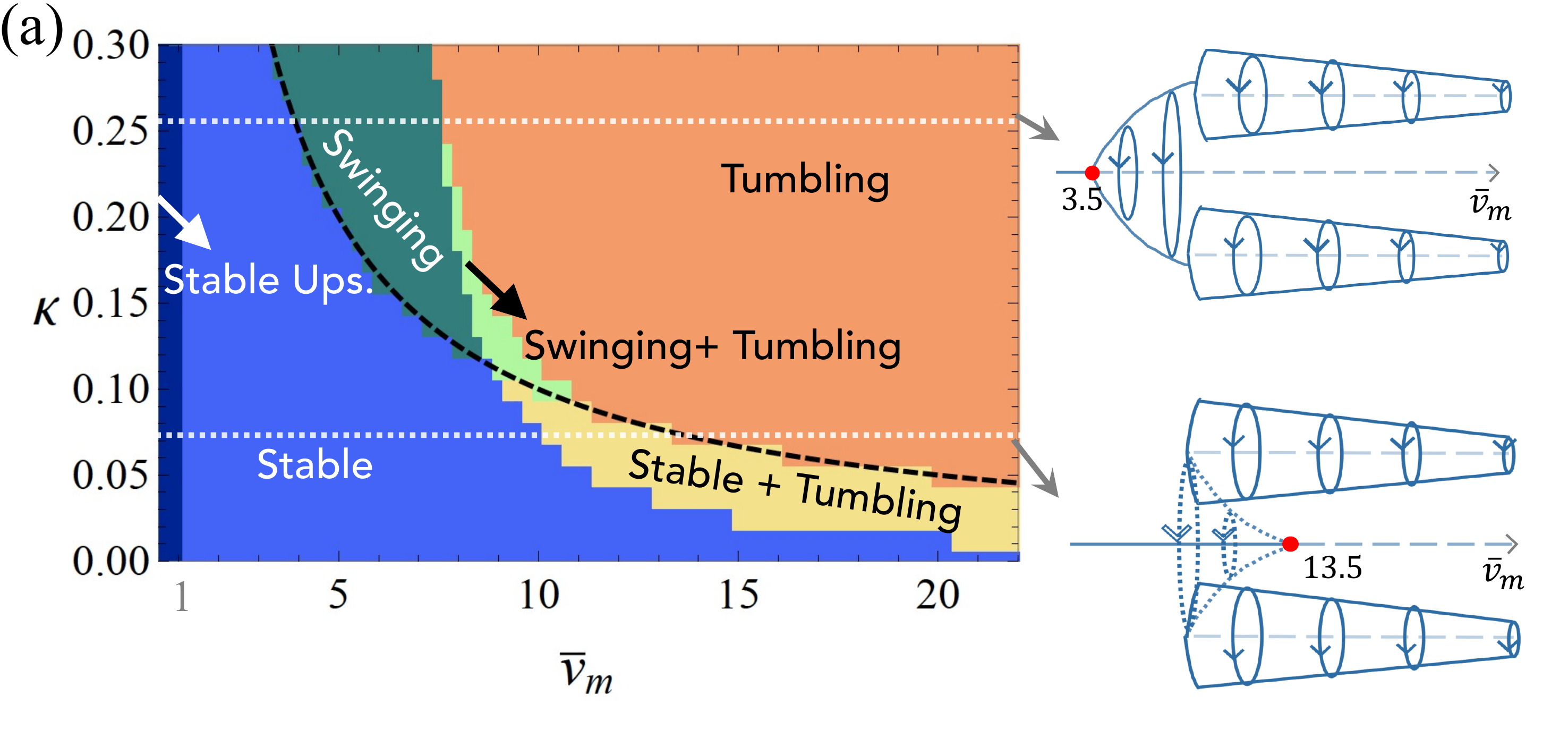}
	\includegraphics[width = 0.99 \columnwidth]{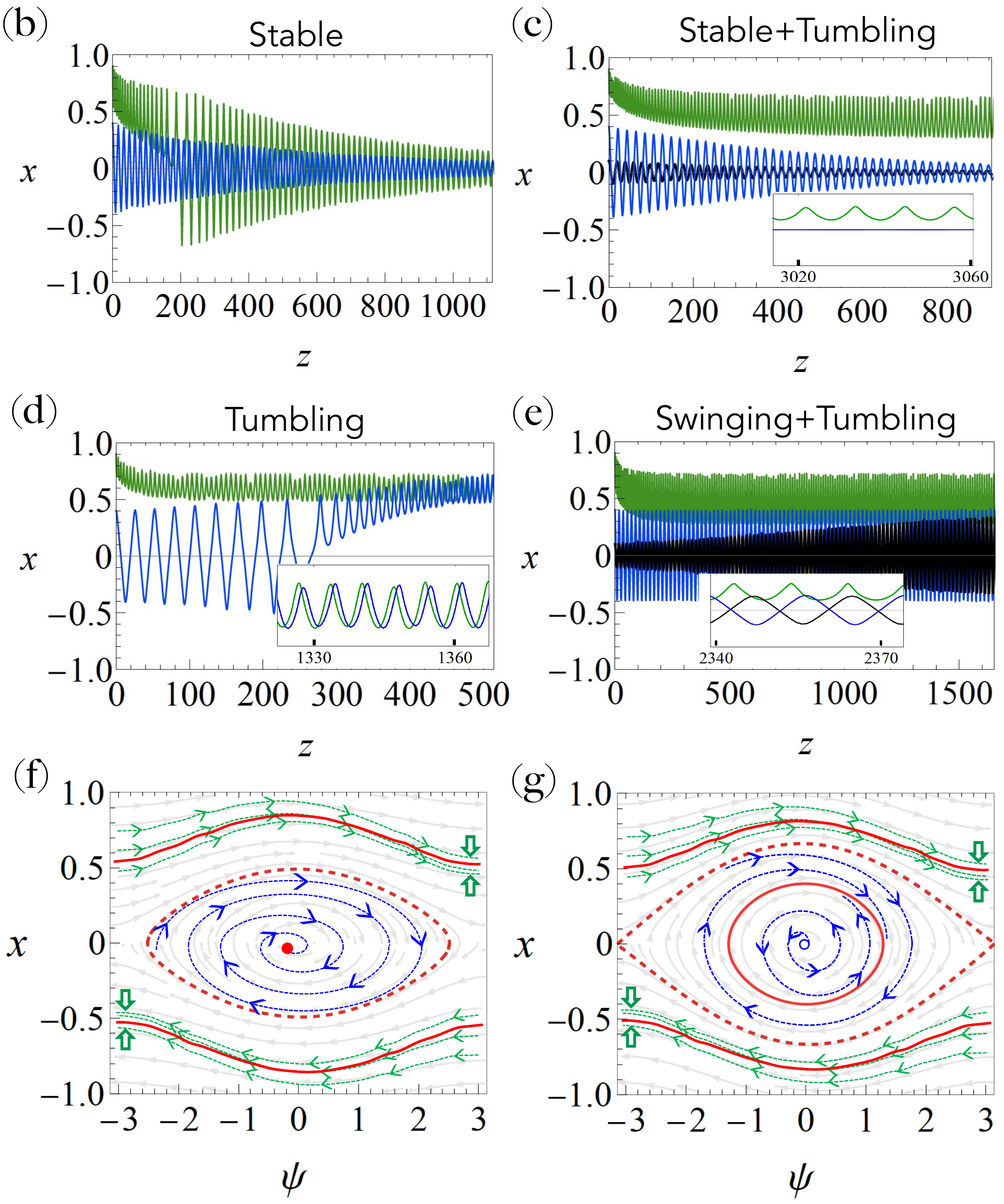}	
		\caption{(a)  
State diagram of a neutral microswimmer. 
To the left  of the black dashed curve, $ \kappa= \bar{v}_{m}^{-1}$, upstream-directed swimming along the centerline is stable.
Along the white dashed lines at $ \kappa=0.25 $ and $ \kappa=0.07$, the bifurcation characteristics are sketched on the right.
Swimmer trajectories for different initial
$x$ positions
[$ x_{0}=0.9$ (green),
$ x_{0}=0.4 $ (blue),
$ x_0 = 0.1 $ (black)]
and parameters:
(b)  $ \bar{v}_{m}=8 $, $ \kappa=0.1 $; (c) $ \bar{v}_{m}=11.5 $, $ \kappa=0.05 $; (d) $ \bar{v}_{m}=14 $, $ \kappa=0.1$;
(e) $ \bar{v}_{m}=9 $, $ \kappa=0.125 $. 
Other parameters are  $ \text{Re}_{p}=0.1 $, and $ x_{eq}=\pm 0.65 $. The insets show zoomed-in trajectories 
in steady state.
(f), (g) Schematic phase portraits for the trajectories in (c) and (e), respectively.
The solid and dashed red lines depict stable and unstable limit cycles, respectively.
	}%
	\label{fig:traj}%
\end{figure}

The inertial lift profile of Eq.\ (\ref{F}) causes a complex dynamics of the microswimmer governed by Eqs. (\ref{kin_dim}),
which we now explore step by step. First of all, we identify two fixed points in the $ x-\psi $ plane at $x=0$, with the
microswimmer either swimming upstream along the centerline ($\psi =0$) or downstream ($\psi = \pm \pi$).
A linear stability analysis reveals the following approximate eigenvalues for these fixed points:
\begin{align}\label{ev}
	\lambda_{1} \approx& \frac{\text{Re}_{p}}{2}  \left( -1+ \kappa \bar{v}_{m} \right) \pm {\rm i} \, \bar{v}_{m}^{1/2}, \nonumber \\
	\lambda_{2} \approx& \frac{\text{Re}_{p}}{2} \left( 1 + \kappa \bar{v}_{m} \right) \pm  \, \bar{v}_{m}^{1/2} . 
\end{align}
Downstream swimming corresponds to a saddle fixed point ($\lambda_2$), while upstream swimming along the centerline ($\lambda_1$) is stable for weak flows ($ \bar{v}_{m} < \kappa^{-1} $) and unstable otherwise. 
The inertial lift profile plotted  in Fig.\ \ref{fig:lift_profile} for 
a moderate flow strength and for different swimmer orientations $\psi$, shows the passive lift velocity at  $\psi = \pi/2$ with an unstable position in the center and the two inertial focusing points at $\pm x_{eq}$.
In the presence of the swimming lift, the centerline position is stabilized at $\psi = 0$. For strong flows ($ \bar{v}_{m} > \kappa^{-1} $)
the centerline position becomes unstable. 
However, the swimmer cannot focus on a non-zero $x$ position, because
due to the non-zero vorticity of the Poiseuille flow\ac{,} it continuously tumbles while drifting downstream. 
In the state diagram presented in Fig.\ \ref{fig:traj}(a), we vary swimmer size $ \kappa $ versus flow strength $\bar{v}_{m}$ and find these two limiting cases in the lower left and upper right region, respectively.
Around the dashed stability line, $\kappa = \bar{v}_{m} ^{-1} $, we observe that fluid inertia engenders rich dynamics, which we discuss now.

We first look at smaller microswimmers with $ \kappa \lesssim 0.1 $ and move along the white dashed line in the state diagram with increasing 
$\bar{v}_m$. At $\bar{v}_m < 1$ the swimmer quickly reaches the centerline and moves upstream, while at moderate flow velocities 
$\bar{v}_m > 1$, it is drifted downstream by the Poiseuille flow and slowly relaxes towards the centerline [see Fig.\ \ref{fig:traj}(b) for
a trajectory in $x$-$z$ plane].  On further increasing $\bar{v}_m$, a subcritical Hopf bifurcation occurs \cite{strogatz}, where the stable centerline 
state  and tumbling motion around $x_{eq}$ coexist [see Fig.\ \ref{fig:traj}(c)]. The schematic phase portrait in Fig.\ \ref{fig:traj}(f)
shows how the stable fixed point and tumbling, a type of stable limit cycle, are separated by an unstable limit cycle. 
According to the bifurcation schematic next to the state diagram,
the unstable limit cycle shrinks to zero and the fixed point becomes unstable. 
Hence, one observes a pure tumbling state [see Fig.\ref{fig:traj}(d)] with an 
amplitude that shrinks with increasing $\bar{v}_m$.

For larger microswimmers we first concentrate on the white dashed line at $\kappa > 0.23$. 
When the fixed point becomes unstable at $\kappa = \bar{v}_m^{-1}$, a supercritical Hopf bifurcation occurs; the stable limit cycle, where the microswimmer performs a swinging motion about the centerline, gradually expands and then splits into two stable tumbling limit cycles. 
However, in the range  $ 0.1 < \kappa < 0.23 $ the swinging limit cycle first enters a small region where it coexists with the tumbling state 
(multiple limit cycles) \cite{perko2013differential} [see Fig.\ \ref{fig:traj}(e)].
They are separated by an unstable limit cycle as the schematic phase portrait in  Fig.\ \ref{fig:traj}(g) shows.
 As the flow rate further increases, the two inner limit cycles annihilate each other and the pure tumbling state remains.

\begin{figure}
	\centering
	\includegraphics[scale=0.37]{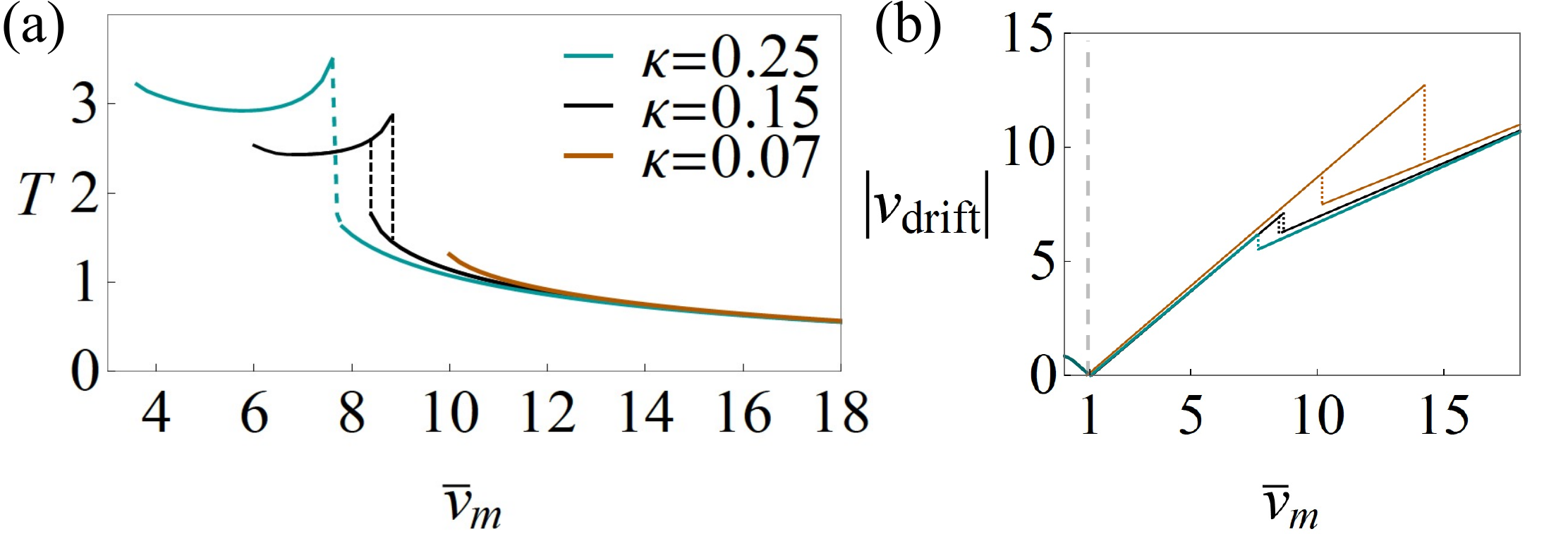}
	\caption{(a) 
	Time period of swinging and tumbling motion and (b) drift speed along the channel axis plotted versus $\bar{v}_{m}$ for different
	$\kappa$. The dashed lines indicate transitions between the two swimming states.
		Time period and drift speed are 
		given in units of $ w/v_{s} $ and $ v_{s} $, respectively. 
	}%
	\label{fig:TP}%
\end{figure} 

In experiments the time period of the swinging and tumbling states as well as the drift velocity of the microswimmer along the channel axis are measurable quantities.
Figure\ \ref{fig:TP}(a) shows the time period $T$ of the oscillatory states exhibited by the source-dipole swimmer for different 
rescaled swimmer sizes $\kappa$.
At $\kappa=0.15$, the two branches have an overlapping $\bar{v}_m$ region. Here, swinging and tumbling states coexist as
indicated in Fig.\ \ref{fig:traj}(a) and
the swimmer state depends on the initial condition.
Dashed lines indicate sharp transitions
between the two states.
As already observed in Fig.\ \ref{fig:traj}(a), larger swimmers enter the 
oscillatory states at lower flow rates.
In the swinging state we obtain a weak dependence of the time period on $\bar{v}_{m}$. Only close to the transition
$T$ rises with $\bar{v}_{m}$ and then, in the tumbling state, it decreases slowly.
Figure\ \ref{fig:TP}(b) shows that the drift speed along the channel axis rises 
linearly with $ \bar{v}_{m}$ with a slope one in the state of centerline swimming as expected. Interestingly,
also in the swinging state ($\kappa=0.15$ and 0.25) the slope is close to one. After the sharp drop to the tumbling state 
indicated by the dashed line, all three curves fall again nearly on top of each other. 
The slope of these straight lines is around 0.5, indicating that tumbling occurs outside of the centerline.

\subsection{Pusher/Puller-type Swimmers}

So far we have concentrated on microswimmers that generate a source-dipole flow field. Since the swimming lift crucially depends on the swimmer's hydrodynamic signature and thus on its propulsion mechanism, we also expect a fundamentally distinct dynamics. Microswimmers that self-propel by rotating or beating flagella, such as \textit{E. coli} and \textit{Chlamydomonas}, generate a force-dipole flow field at the leading order \cite{pedley1992hydrodynamic,berke2008hydrodynamic}: 
$\IB{v}_0 = \mathcal{P}  \IB{r} \big[   \frac{-1}{r^{3}} + 3 \frac{\left(\IB{r} \cdot \IB{p}  \right)^{2} }{r^{5 }}  \big] $.
Here $ \mathcal{P} $ is the dimensionless force-dipole strength normalized by 
$ 8\pi \mu a^{2} v_s $, which depends on the swimming mechanism \cite{berke2008hydrodynamic,drescher2010direct,drescher2011fluid}.
Earlier studies on \textit{E. coli} \cite{drescher2011fluid,chattopadhyay2006swimming,berke2008hydrodynamic} and \textit{Chlamydomonas} \cite{minoura1995strikingly} suggest that $ |\mathcal{P}| $ varies roughly between 0.04 - 0.3.

The slower decay of the force-dipole field compared to the source dipole means that the swimmer lift  scales with $\text{Re}_{p}$ as
an order-of-magnitude analysis reveals (see supplemental material).
Thus, similar to the case of passive inertial lift \cite{ho1974,asmolov1999inertial}, one can use either regular perturbation theory or matched asymptotic expansions to calculate
the swimming lift in leading order of $\text{Re}_p$
\footnote{A comparison of results 
from singular perturbation approach 
of \citet{asmolov1999inertial} and results 
using regular perturbation 
theory
\cite{ho1974}, which strictly
requires a channel Reynolds number $\text{Re}_{c} \ll 1$,
shows a close match of the lift-force profiles at $ \text{Re}_{c}=15 $ (see Fig. 8 in \cite{asmolov1999inertial}). 
This suggests a smooth transition between the two approaches.}.
Thus, we continue with the approach used for neutral microswimmers and employ regular perturbation theory in combination with the reciprocal theorem in Eq.\ (\ref{lift}), as detailed in the supplemental material. 
The slower decay of the force-dipole field poses an additional challenge: one has to account for the finite integration domain of the microchannel,
otherwise the lift would  diverge logarithmically.
Thus, we correct the zeroth-order flow field $\IB{v}_0$ by including wall 
terms, which we obtain from the method of reflections. Our investigation shows that the angular dependence of the force-dipole swimming lift, 
$\mathcal{F}_{\text{swim}} \propto  \sin 2 \psi$, differs from that of the source dipole.
Fitting the numerical results for $\mathcal{F}_{\text{swim}}$, we can approximate the total inertial lift velocity in units of $v_s$ by
\begin{equation*}
\label{dynamic_FD}
\mathcal{F} =  \text{Re}_{p}  \left[ \kappa \, \bar{v}_{m}  x \left( 1-\frac{x^{2}}{x_{eq}^{2}} \right)  +  \mathcal{P} \left(1-2x^{2}\right) \sin 2\psi /2  \right].
\end{equation*}

\begin{figure}
	\centering
	\includegraphics[width= 1 \columnwidth]{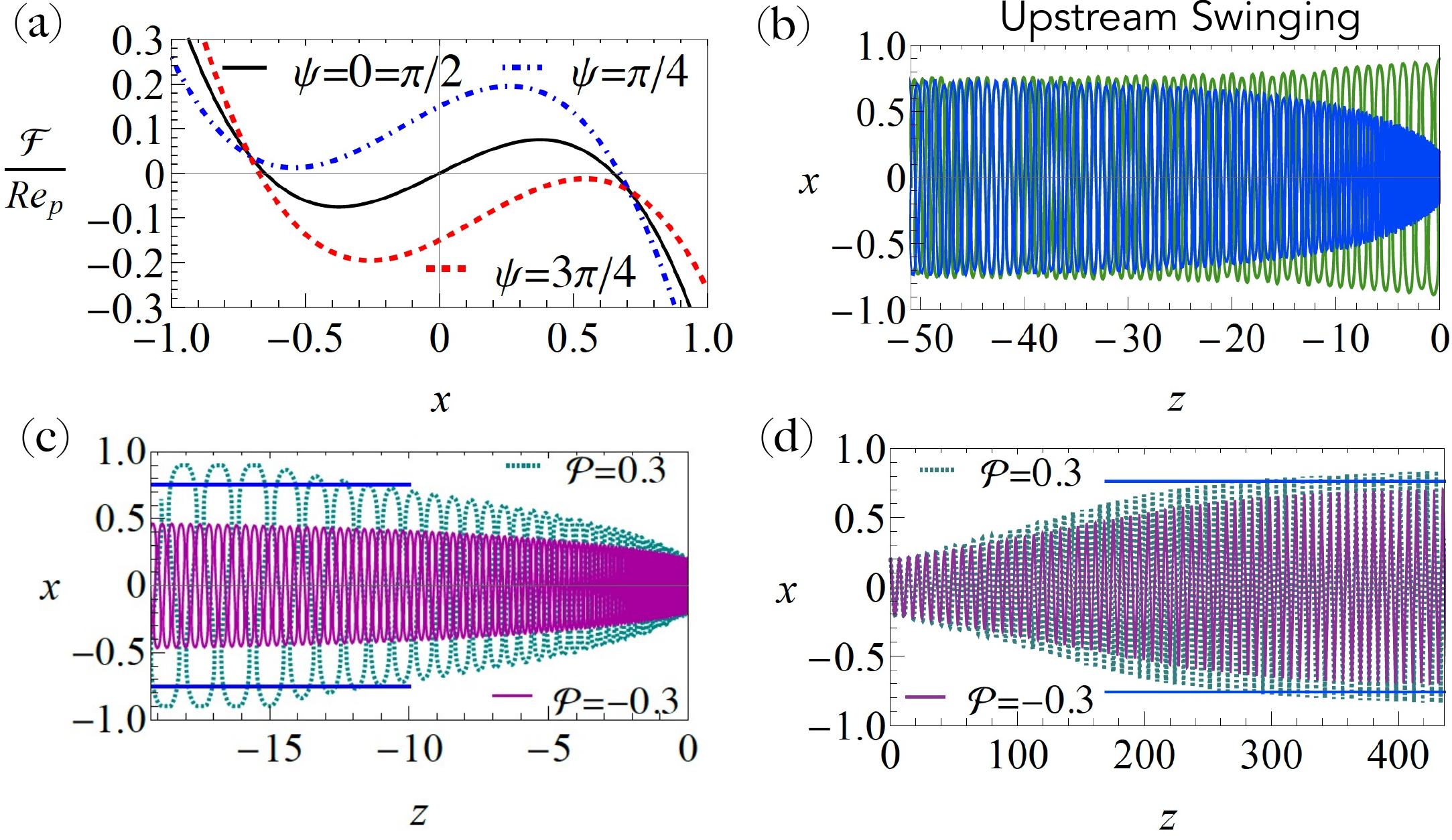}
	\caption{  (a) Inertial lift-velocity profile of a pusher ($ \mathcal{P}=0.3 $) with
		$ \kappa =0.1 $, $ x_{eq}= \pm 0.65 $, and $ \bar{v}_{m}=3 $. 
		(b) Upstream swinging trajectories for $ \bar{v}_{m} = 1$.
		The bottom row shows hydrodynamics wall effects on the upstream and downstream motion of  a pusher  ($ \mathcal{P}=0.3 $) 	and puller ($ \mathcal{P}=-0.3 $): (c) $ \bar{v}_{m} = 1 $, (d) $ \bar{v}_{m}=3 $.
		The solid blue line depicts the limit cycle amplitude 
		of (b). 
		$\text{Re}_{p} = 0.1$ is used in all figures.
	}
	\label{fig:traj_FD}%
\end{figure}

In Fig.\ \ref{fig:traj_FD}(a), the lift-velocity profile for a force dipole shows a clear difference to the profile in Fig.\ \ref{fig:lift_profile}. Compared to the passive lift ($\psi = 0,\pi/2$, and $\psi$), the profile either shifts up or down for varying $\psi$. 
Thus, depending on $\mathcal{P}$ and $ \bar{v}_{m} \kappa $,
the fixed point ($\mathcal{F} = 0$) in one channel half can vanish completely. 
We note that the profiles of force dipoles with the same strength but opposite signs follow from each other by
adding $\pi/2$ to $\psi$.

Although the fixed points are identical to the previous case, the stability analysis 
with the eigenvalues
\begin{equation}
\lambda_{1} \approx \frac{\text{Re}_{p}}{2} \, \kappa \bar{v}_{m}  \pm {\rm i} \, \bar{v}_{m}^{1/2} \quad \text{and} \quad
\lambda_{2} \approx \frac{\text{Re}_{p}}{2}  \, \kappa \bar{v}_{m}  \pm \bar{v}_{m}^{1/2} .\nonumber
\end{equation}
reveals that upstream swimming ($\psi = 0$) is always unstable, as suggested by the lift velocity. Through an unstable spiral, the trajectories enter a stable limit cycle, which for lower flow rates corresponds to a swinging motion about the centerline.
The swimmer effectively swims upstream for $ \bar{v}_{m}<1$ as depicted in Fig. \ref{fig:traj_FD}(b), while it moves 
downstream for $ \bar{v}_{m}>1$, similar to the black trajectory in Fig.\ \ref{fig:traj}(e).

Hydrodynamic wall interactions of the force-dipole field add weak modifications of the order of $\kappa^{2}$ and $\kappa^{3}$ to the
evolution equations of position and orientation, respectively \cite{zottl2012nonlinear,kim2013,ibrahim2016walls,shaik2017motion}.
Therefore, they mainly influence
the dynamics when the flow rates are weak, \emph{i.e.}, for upstream swinging motion.
Figure\ \ref{fig:traj_FD}(c) shows  a pusher approaching the wall as the hydrodynamic interactions are attractive \cite{berke2008hydrodynamic}. 
Since the strong vorticity near the walls re-orients the swimmer,
it will ultimately oscillate between both walls.
In contrast, pullers are hydrodynamically repelled from walls  \cite{zottl2012nonlinear}
and hence swim in a swinging limit cycle with an amplitude smaller compared to Fig.\ \ref{fig:traj_FD}(b).
Finally, Fig.\ \ref{fig:traj_FD}(d) shows that 
downstream swinging in stronger flows is hardly affected.
For neutral swimmers the wall effects are weaker by an additional factor of  $\kappa$ \cite{spagnolie_lauga_2012,shaik2017motion,ibrahim2016walls,choudhary2021self} and we verified that they do not have a significant 
effect on the dynamics.

\begin{figure}
	\centering
	\includegraphics[scale=0.385]{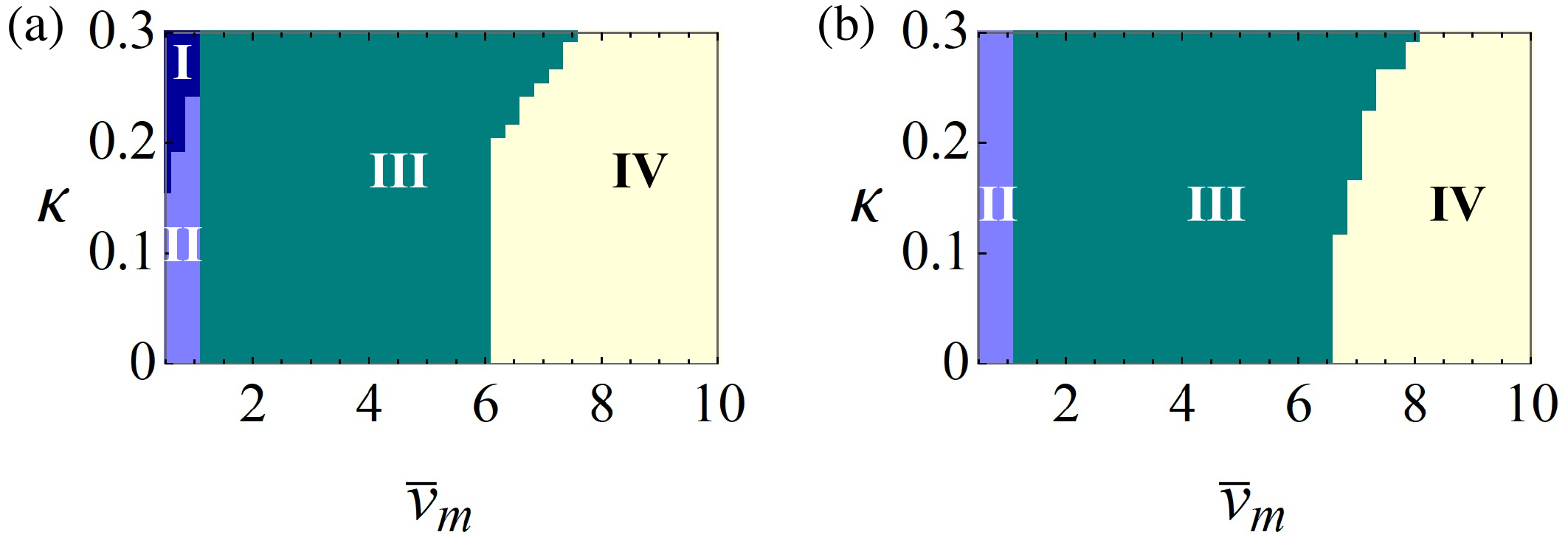}
	\caption{State diagram,
	particle size $\kappa$ versus flow speed $\bar{v}_m$, for (a) puller ($ \mathcal{P}=-0.3 $) 
	and (b) Pusher ($ \mathcal{P}=0.3 $) at $\text{Re}_{p} = 0.1$.
Regions I: centerline upstream swimming, II: upstream swinging, III: downstream swinging, and IV: tumbling.}%
	\label{fig:Phase_FD}%
\end{figure}

In Figs.\ \ \ref{fig:Phase_FD}(a) and (b) we show the resulting state diagrams for a puller and pusher, respectively. The diagrams 
are clearly disparate to that of a
neutral swimmer [fig.\ \ref{fig:traj}(a)].
For flow rates $\bar{v}_m$ below one, larger pullers swim upstream along the centerline (region I) since hydrodynamic wall interactions 
dominate the inertial lift and push pullers to the center. 
Otherwise, pushers and puller show upstream swinging (region II)
and for $\bar{v}_m > 1$ downstream swinging (region III). At even larger $\bar{v}_m$ they transition into the tumbling state (region IV).
For pushers, this transition occurs at larger $\bar{v}_m$ due to the hydrodynamic wall interactions.
Finally, in the supplemental material we provide the time period of the oscillatory states and the axial drift speed
as a function of $\bar{v}_m$ for pusher and puller with $ \mathcal{P}= \pm 0.3$.

\section{Conclusions and Outlook}

In summary, we have studied how swimming at low fluid inertia in Poiseuille flow adds a swimming lift to the known passive 
inertial lift velocity. We have concentrated on the generic source-dipole and force-dipole microswimmers and showed that their 
swimming lift velocities depend differently on the lateral swimmer position and orientation. This gives rise to the emergence of 
novel complex dynamics including bistable states, where tumbling coexists with stable centerline swimming or swinging. The 
Reynolds number determines the overall dynamics relative to the flow speed. Deriving a non-linear oscillator equation for $\psi$ 
in full analogy to Ref.\ \cite{zottl2012nonlinear}, reveals a reduced relaxation time $\propto \text{Re}_p^{-1}$ towards the 
stationary states.

Recent experimental studies \cite{rusconi2014bacterial,drescher2010direct,barry2015shear,tung2015emergence} operate within the 
parameter ranges of microswimmer size, $ 10-200 \, \mu $m, and channel width, $ 100-500 \,  \mu $m. Thus for the maximum flow speed 
$ v_{m} \sim 1 \,  $mm/s, $ \text{Re}_{p} $ ranges from $0.001 - 0.1 $ and the time taken to attain steady states, 
$ {w}/{v_{s} \text{Re}_{p}} $, roughly varies from $10-10^{3} \, $s for narrow microchannels. These estimates suggest
that effects of fluid inertia are observable for large microswimmers ($ \gtrsim 50 \mu $m) and moderate to strong flows.
For instance, \textit{Volvox carteri} will be of interest as it has a radius of $ \sim 200 \mu m $ and swims with $ \sim 200 \mu m/s $ \cite{drescher2010direct}.
Additionally, artificial microswimmers with tunable high speeds larger than $200 \mu\text{m/s}$ exist \cite{ren20193d,aghakhani2020acoustically,zhou2021magnetically}. 
All this should offer the possibility to experimentally observe the dynamic features reported here at small but non-negligible fluid 
inertia depending on the hydrodynamic signature of a microswimmer.
Furthermore,
 the current insights may 
encourage investigations in marine ecosystem, where recent literature \cite{woodward2019physical,beron2021nonlinear} suggests that 
inertial lift can drive planktons out of the turbulent eddies and induce plankton blooms.

Our work extends the research on microswimmers into a new direction by bringing the role of fluid inertia into focus, which
has not been looked at so far. For passive particles this has spawned the field of inertial microfluidics\ \cite{di2009inertial,zhang2016}. 
We envisage a similar development for microswimmers, which offers novel aspects to look at. For example, 
elongated microswimmers perform Jeffery orbits \cite{jeffery1922motion}, which also influence
their dynamics in a Poiseuille flow \cite{zottl2013periodic}. Adding them to the current work is not straightforward since fluid inertia induces an orientational drift \cite{einarsson2015rotation}.
The hydrodynamics of the swimming motion might  also add an active component to the Jeffery orbits.
We finally note that thermal or biological noise acting
on the swimmer orientation will disturb the motion in the limit cycles and also induce transitions between coexisting states but not influence the principal behavior outlined in this article.

\vspace{4mm}

\begin{acknowledgments}
Support from the Alexander von Humboldt Foundation is gratefully acknowledged.
\end{acknowledgments}

	\bibliography{Akash}

\end{document}